\documentclass[lettersize,journal]{IEEEtran}
\usepackage{amsmath,amsfonts}
\usepackage{algorithmic}
\usepackage{array}
\usepackage[caption=false,font=normalsize,labelfont=sf,textfont=sf]{subfig}
\usepackage{textcomp}
\usepackage{stfloats}
\usepackage{url}
\usepackage{verbatim}
\usepackage{graphicx}
\def\BibTeX{{\rm B\kern-.05em{\sc i\kern-.025em b}\kern-.08em
    T\kern-.1667em\lower.7ex\hbox{E}\kern-.125emX}}
\usepackage{balance}

\usepackage{booktabs}                        
\usepackage{subcaption}   
\usepackage{color}
\usepackage{amsmath}
\usepackage{amssymb}
\usepackage{graphicx}
 \usepackage{dsfont}
\usepackage{amsfonts}
\usepackage{amsthm} 
\usepackage{stmaryrd}
\usepackage[all]{xy}
\usepackage{multirow}

\usepackage{qcircuit}
 
\def\>{\ensuremath{\rangle}}
\def\<{\ensuremath{\langle}}

\newcommand{\hs}{\mathcal{H}}

\newtheorem{defn}{Definition}[section]

\newtheorem{exam}{Example}[section]

\newtheorem{rem}{Remark}[section]

\begin{document}
\title{Quantum Recursive Programming with Quantum Case Statements}
\author{Mingsheng Ying and Zhicheng Zhang 
\thanks{Mingsheng Ying is with the State Key Laboratory of Computer Science, Institute of Software, Chinese Academy of Sciences, and the Department of Computer Science and Technology, Tsinghua University, China. E-mail: yingms@ios.ac.cn; yingmsh@tsinghua.edu.cn.}
\thanks{Zhicheng Zhang is with the Centre for Quantum Software and Information, University of Technology Sydney, Australia. E-mail: Zhicheng.Zhang@student.uts.edu.au.}
}

\markboth{Journal of \LaTeX\ Class Files,~Vol.~18, No.~9, September~2020}
{How to Use the IEEEtran \LaTeX \ Templates}

\maketitle

\begin{abstract} We introduce a novel scheme of quantum recursive programming, in which large unitary transformations, i.e. quantum gates, can be recursively defined using quantum case statements, which are quantum counterparts of conditionals and case statements extensively used in classical programming. A simple programming language for supporting this kind of quantum recursion is defined, and its semantics is formally described. A series of examples are presented to show that some quantum algorithms can be elegantly written as quantum recursive programs. 
\end{abstract}


\section{Introduction}\label{sec-intro}
\IEEEPARstart{T}{he} computer programming pioneers like Dijkstra, Hoare and many others had persuaded a high level of elegance in early programming research by introducing effective program constructs and programming schemes. 
In particular, iteration and recursion were employed to describe repetitive tasks without requiring a large number of steps to be specified individually.    
Typical examples include:  (i) Quicksort can be elegantly expressed as a recursive program \cite{Hoare62}; and (ii) Euclid's algorithm that computes the greatest common divisor (gcd) of two positive integers can be elegantly written as a \textbf{do}-loop:
\begin{equation}\label{Euclid}\begin{split}&\mathbf{do}\ x>y\rightarrow x:=x-y\\
&\square\ \ x<y\rightarrow y:=y-x\\
&\mathbf{od}
\end{split}\end{equation} in the guarded commands language (GCL) \cite{Dijk75}. 

How can we achieve the same level of elegance in quantum programming? Indeed, at this moment, the majority of quantum programming research focuses on relatively low-level features, and the higher-level elegance of quantum programming has not been seriously considered at all. This \textit{short paper} presents an attempt toward the elegance in quantum programming by introducing a novel scheme of quantum recursion. 

As is well-known, \textbf{if-then-else} conditionals, or more general case statements, are extensively used in recursive definitions of functions in classical programming. An example is the program (\ref{Euclid}) of the Euclid's algorithm. It has been realised that two fundamentally different kinds of case statements can be defined in quantum programming \cite{Ying16}:
\begin{enumerate}\item \textit{Measurement-based case statements} are usually given in the following form: 
\begin{equation}\label{qcase-1}\mathbf{if}\ (\square i\cdot M[q]=m_i\rightarrow P_i)\ \mathbf{fi}\end{equation}
where $q$ is a quantum variable and $M$ a measurement performed on $q$ with possible outcomes $m_i$'s, and for each $i$, $P_i$ is a subprogram. The statement (\ref{qcase-1}) selects a command according to the outcome of measurement $M$: if the outcome is $m_i$, then the corresponding command $P_i$ will be executed. It is worth noting that the control flow of (\ref{qcase-1}) is classical because the selection of commands in it is based on classical information --- the outcomes of a quantum measurement.
\item \textit{Quantum case statements} are usually defined using a quantum \textquotedblleft coin\textquotedblright\ in the following form:
\begin{equation}\label{qcase-2}\mathbf{qif} [c]\ (\square i\cdot |i\rangle\rightarrow P_i)\ \mathbf{fiq}\end{equation} 
where $\{|i\rangle\}$ is an orthonormal basis of the state Hilbert space of an \textit{external} \textquotedblleft coin\textquotedblright\ system $c$, and the selection of subprograms $P_i$'s is made according to the basis states $|i\rangle$ of the \textquotedblleft coin\textquotedblright\ space. A fundamental difference between (\ref{qcase-1}) and (\ref{qcase-2}) is that the control flow of (\ref{qcase-2}) is quantum because the basis states of quantum \textquotedblleft coin\textquotedblright\ $c$ can be superposed and thus $c$ carries quantum information rather than classical information (for more about quantum control flow, see \cite{Ying16}, Chapter 6 and \cite{Yuan}).\end{enumerate}
The scheme of recursion with measurement-based case statements (\ref{qcase-1}) has already been studied in the literature, and was termed in \cite{Ying16} as \textit{recursive quantum programming} for the recursion is executed along classical control flow.
In this paper, we consider a new scheme of recursion with quantum case statements (\ref{qcase-2}). 
An important difference between this scheme of quantum recursion and the previous one is as follows: in this scheme, procedure identifiers can occur in different branches of a quantum case statement of the form (\ref{qcase-2}) and thus recursive calls to them may happen in the way of quantum parallelism as a superposition of execution paths. Thus, we call the new scheme \textit{quantum recursive programming} for the execution is executed along quantum control flow.  
As will be shown in a series of examples, an important class of large quantum gates can be defined and quantum algorithms can be described in the new scheme of quantum recursion conveniently and elegantly.

This paper is organised as follows. As a basis for defining quantum recursion, we introduce quantum arrays in Section \ref{sec-arrays}. Our quantum recursive programs are then introduced in several steps. We start from defining a quantum circuit description language $\mathbf{QC}$ in Section \ref{sec-circuits}. In Section \ref{sec-c-variables}, $\mathbf{QC}$ is embedded into a simple classical programming language. Then quantum recursive programs without parameters and their semantics are defined in Section \ref{sec-non-parameters}. The quantum recursive programs considered in \ref{sec-non-parameters} are generalised by equipped with parameters in Section \ref{sec-parameters}, where actual parameters are described using the classical programming language presented in Section \ref{sec-c-variables}.  

\section{Quantum Arrays}\label{sec-arrays}

In this section, we introduce the notions of quantum array and subscripted quantum variable, which will be needed in defining quantum recursive programs.    

\subsection{Quantum types} As a basis, let us first define the notion of quantum type. Recall that a basic classical type $T$ denotes an intended set of values. Similarly, a basic quantum type $\hs$ denotes an intended Hilbert space. It will be considered as the state space of a simple quantum variable. We will also use higher quantum type of the form: 
\begin{equation}\label{h-type}T_1\times ...\times T_n\rightarrow\hs\end{equation}
where $T_1,...,T_n$ are basic classical types, and $\hs$ is a basic quantum type. Mathematically, this type denotes the following tensor power of Hilbert space $\hs$ (i.e. tensor product of multiple copies of $\hs$):
\begin{equation}\label{h-space}\hs^{\otimes (T_1\times ...\times T_n)}=\bigotimes_{t_1\in T_1,...,t_n\in T_n}\hs_{t_1,...,t_n}\end{equation} where $\hs_{t_1,...,t_n}=\hs$ for all $t_1\in T_1,...,t_n\in T_n$. 
Intuitively, if $\hs$ is the state space of a quantum system $A$, then according to the basic postulates of quantum mechanics, the Hilbert space (\ref{h-space}) is the state space of the composite system consisting of those quantum systems indexed by $(t_1,...,t_n)\in T_1\times ...\times T_n$, each of which is identical to $A$. 

A notable basic difference between higher classical types and quantum ones is that some states in the space (\ref{h-space}) are entangled between the quantum systems indexed by different $(t_1,...,t_n)\in T_1\times ...\times T_n$.   

\begin{exam}Let $\hs_2$ be the qubit type denoting the $2$-dimensional Hilbert space. If $q$ is a qubit array of type $\mathbf{integer}\rightarrow\hs_2$, where $\mathbf{integer}$ is the (classical) integer type, then for any two integers $k\leq l$, section $q[k:l]$ stands for the restriction of $q$ to the interval $[k:l]=\{{\rm integer}\ i|k\leq i\leq l\}$. For example,
$$\frac{\bigotimes_{i=k}^l|0\rangle_i+\bigotimes_{i=k}^l|1\rangle_i}{\sqrt{2}}$$ is an entangled state of the qubits labelled $k$ through $l$. 
\end{exam}

\subsection{Quantum variables}In this paper, we will use two sorts of quantum variables:\begin{itemize}\item simple quantum variables, of a basic quantum type, say $\hs$; \item array quantum variables, of a higher quantum type, say $T_1\times ...\times T_n\rightarrow\hs$.  
\end{itemize}

\begin{defn}Let $q$ be an array quantum variable of the type $T_1\times ...\times T_n\rightarrow\hs$, and for each $1\leq i\leq n$, let $s_i$ be a classical expression of type $T_i$. Then $q[s_1,...,s_n]$ is called a subscripted quantum variable of type $\hs$.\end{defn}

Intuitively, array variable $q$ denotes a quantum system composed of subsystems indexed by $(t_1,...,t_n)\in T_1\times ...\times T_n$. Thus, whenever expression $s_i$ is evaluated to a value $t_i\in T_i$ for each $i$, then $q[s_1,...,s_n]$ indicates the system of index tuple $(t_1,...,t_n)$. For example, let $q$ be a qubit array of type $\mathbf{integer}\times\mathbf{integer}\rightarrow\hs_2$. Then $q[2x+y,7-3y]$ is a subscripted qubit variable; in particular, if $x=5$ and $y=-1$ in the current classical state, then it stands for the qubit $q[9,10]$.

\section{A Quantum Circuit Description Language}\label{sec-circuits}

In this section, we introduce a quantum circuit description language $\mathbf{QC}$. The major difference between $\mathbf{QC}$ and other languages in the previous literature for the same purpose is that the construct of quantum case statement is added into $\mathbf{QC}$. As pointed out in Section \ref{sec-intro}, this language will be expanded gradually in the subsequent sections for recursive definition of large quantum gates and algorithms.  

\subsection{Syntax}We assume that the alphabet of \textbf{QC} consists of: \begin{itemize}\item A set $\mathit{QV}$ of simple or subscripted quantum variables;
\item A set $\mathcal{U}$ of unitary matrix constants.   
\end{itemize}
The unitary matrix constants in $\mathcal{U}$ will be instantiated in practical applications. We fix the following notations:
\begin{itemize}\item[-] As defined in Section \ref{sec-arrays}, each quantum variable $q\in\mathit{QV}$ assumes a type $T(q)$. This means that variable $q$ stands for a quantum system with the Hilbert space denoted by $T(q)$ as its state space.
\item A sequence $\overline{q}=q_1,...,q_n$ of distinct quantum variables is called a quantum register. It denotes a composite quantum system consisting of subsystems $q_1,...,q_n$. Its type is defined as the tensor product $T(\overline{q})=T(q_1)\otimes ...\otimes T(q_n)$ of the types of $q_1,...,q_n$. 
For simplicity of the presentation, we often identify $\overline{q}$ with the set $\{q_1,...,q_n\}$ of quantum variables occurring in $\overline{q}$.   
\item[-] Each unitary matrix constant $U\in\mathcal{U}$ assumes a type of the form $T(U)=\hs_1\otimes ...\otimes\hs_n$. This means that the unitary transformation denoted by $U$ can be performed on a composite quantum system consisting of $n$ subsystems with types $\hs_1,...,\hs_2$, respectively. Thus, if $\overline{q}=q_1,...,q_n$ is a quantum register with $T(q_i)=\hs_i$ for $i=1,...,n$; that is, the types of $U$ and register $\overline{q}$ match, then $U[\overline{q}]$ can be thought of as a quantum gate with quantum wires $q_1,...,q_n$.  
\end{itemize}

\begin{defn}\label{def-circ}Quantum circuits $C\in\mathbf{QC}$ are defined by the syntax:
\begin{equation}\label{def-syntax}\begin{split}C::=\ U[\overline{q}]\ &|\ C_1;C_2\\ &|\ \mathbf{qif}[\overline{q}]\ \left(\square_{i=1}^d |\psi_i\rangle\rightarrow C_i\right)\ \mathbf{fiq}\end{split}\end{equation}
More precisely, they are inductively defined by the following clauses, where $\mathit{qv}(C)$ denotes the quantum variables in $C$:
\begin{enumerate} 
\item[(1)] \textbf{Basic gates}: If $U\in\mathcal{U}$ is a unitary matrix constant and $\overline{q}$ is a quantum register such that their types match, then quantum gate $U[\overline{q}]$ is a circuit, and $\mathit{qv}(U[\overline{q}])=\overline{q}$, ;  
\item[(2)] \textbf{Sequential composition}: If $C_1$ and $C_2$ are circuits, then $C\equiv C_1;C_2$ is a circuit too, and $\mathit{qv}(C)=\mathit{qv}(C_1)\cup\mathit{qv}(C_2);$
\item[(3)] \textbf{Quantum case statement}: If $\overline{q}$ is a quantum register, $\left\{|\psi_i\rangle\right\}_{i=1}^d$ is an orthonormal basis of the Hilbert space denoted by the type $T(\overline{q})$, and $C_i$ $(i=1,...,d)$ are circuits with \begin{equation}\label{coin-cond}\overline{q}\cap\left(\bigcup_{i=1}^d\mathit{qv}(C_i)\right)=\emptyset,\end{equation} then \begin{equation}\label{q-mux}C\equiv \mathbf{qif}[\overline{q}]\ \left(\square_{i=1}^d |\psi_i\rangle\rightarrow C_i\right)\ \mathbf{fiq}\end{equation} is a circuit, and $\mathit{qv}(C)=\overline{q}\cup\left(\bigcup_{i=1}^d\mathit{qv}(C_i)\right).$
\end{enumerate}\end{defn}

Intuitively, each $C\in\mathbf{QC}$ represents a circuit with quantum wires $\mathit{qv}(C)$. The circuit constructs introduced in the above definition are explained as follows.  
\begin{enumerate}\item[(i)] The circuit $C_1;C_2$ in clause (2) stands for the sequential composition of circuits $C_1$ and $C_2$. Indeed, if $C_1$ and $C_2$ do not share quantum variables; that is, $\mathit{qv}(C_1)\cap \mathit{qv}(C_2)=\emptyset$, we can also define their parallel composition $C_1\otimes C_2$. But $C_1;C_2$ and $C_1\otimes C_2$ are semantically equivalent whenever $\mathit{qv}(C_1)\cap \mathit{qv}(C_2)=\emptyset$. So, the parallel composition is not included in the above definition.      
\item [(ii)] The quantum case statement defined in equation (\ref{q-mux}) is a straightforward generalisation of (\ref{qcase-2}) with a quantum \textquotedblleft coin\textquotedblright\ $c$ being replaced by a sequence $\overline{q}$ of quantum variables. The condition (\ref{coin-cond}) means that $\overline{q}$ is a system external to all $C_i$ $(i=1,...,d)$. Semantically, quantum case statement (\ref{q-mux}) is a quantum multiplexor (i.e. a multi-way generalisation of conditional) \cite{Markov}. A multiplexor can be understood as a switch that passes one of its data inputs through to the output, as a function of a set of select inputs. Here, $\overline{q}$ is the select register, and if $\overline{q}$ is in state $|\psi_i\rangle$, then a quantum datum $|\varphi\rangle$ is inputted to the corresponding circuit $C_i$ and an output $|\varphi_i\rangle$ is obtained at the end of $C_i$. A basic difference between classical and quantum multiplexors is that the quantum select register $q$ can be in a superposition of $|\psi_i\rangle$ $(i=1,...,d)$, say $\sum_{i=1}^d\alpha_i|\psi_i\rangle$. In this case, the output is then $\sum_{i=1}^d\alpha_i|\varphi_i\rangle$, a superposition of the outputs of different circuits $C_i$ $(i=1,...,d)$. This point will be seen more clearly from the operational semantics defined below. 
\end{enumerate}

\begin{rem}A much more general notion of quantum case statement in which quantum measurements may occur was introduced in \cite{Ying16}, Chapter 6. In this paper, however, the notion of quantum case statement is restricted to quantum circuits (thus, unitary transformations) so that its semantics can be more clearly defined. 
\end{rem}

\subsection{Operational semantics}

For each quantum variable $q\in\mathit{QV}$, assume its type $T(q)$ denotes Hilbert space $\hs_q$. Then for any set $X\subseteq\mathit{QV}$ of quantum variables, the state space of the quantum variables in $X$ is the tensor product    
$\hs_X=\bigotimes_{q\in X}\hs_q.$ In particular, the state space of all quantum variables is $\hs_\mathit{QV}$. 

A configuration is defined as a pair $(C,|\psi\rangle)$, where $C\in\mathbf{QC}$ is a quantum circuit or $C=\ \downarrow$ stands for termination, and $|\psi\rangle$ is a pure quantum state in Hilbert space $\hs_X$ for some $\mathit{qv}(C)\subseteq X\subseteq\mathit{QV}$. We write $\mathcal{C}$ for the set of all configurations with circuits in $\mathbf{QC}$. 

\begin{defn}The operational semantics of quantum circuits in $\mathbf{QC}$ is the transition relation $\rightarrow\ \subseteq\mathcal{C}\times\mathcal{C}$ between configurations that is defined by the transitional rules given in Table \ref{circuit-rules}.
\begin{table*}[t]
\begin{equation*}\begin{split}&%({\rm SK})\ \ 
%\rightarrow (\downarrow,|\psi\rangle)\qquad\qquad\qquad\qquad
({\rm GA})\ \ 
(U[\overline{q}],|\psi\rangle)\rightarrow \left(
\downarrow,(U\otimes I_{\mathit{QV}\setminus\overline{q}})|\psi\rangle\right)\qquad\qquad\qquad\qquad
({\rm SC})\ \ \frac{(C_1,|\psi\rangle)\rightarrow (
C_1^{\prime},|\psi^{\prime}\rangle)}{(
C_1;C_2,|\psi\rangle)\rightarrow (C_1^{\prime};C_2,|\psi^\prime\rangle)}\\
&({\rm QC})\ \ \frac{|\psi\rangle=\sum_{i=1}^d\alpha_i|\psi_i\rangle_{\overline{q}}|\theta_i\rangle\qquad\qquad (C_i,|\theta_i\rangle)\rightarrow^\ast(\downarrow,|\theta_i^\prime\rangle)\ (i=1,...,n)}{\left(\mathbf{qif}[\overline{q}]\ \left(\square_{i=1}^d |\psi_i\rangle\rightarrow C_i\right)\ \mathbf{fiq},|\psi\rangle\right)\rightarrow \left(\downarrow,\sum_{i=1}^d\alpha_i|\psi_i\rangle_{\overline{q}}|\theta_i^\prime\rangle\right)}
\end{split}\end{equation*}
\caption{Transition Rules for Quantum Circuits. In rule (GA), $I_{\mathit{QV}\setminus\overline{q}}$ stands for the identity operator on the Hilbert space $\hs_{\mathit{QV}\setminus\overline{q}}$. In rule (SC), we make the convention 
that $\downarrow;C_2=C_2.$ In rule (QC), $\rightarrow^\ast$ denotes the reflexive and transitive closure of relation $\rightarrow$.}\label{circuit-rules}
\end{table*}
\end{defn}

The transition rules (GA) and (SC) are self-explanatory. But the rule (QC) needs an careful explanation. First, for any state $|\psi\rangle$ in a Hilbert space $\hs_X$ with $X\supseteq\overline{q}$, we can always write it in the form of $|\psi\rangle=\sum_{i=1}^d\alpha_i|\psi_i\rangle_{\overline{q}}|\theta_i\rangle,$ as done in the premise of the rule (QC), because $\hs_X=\hs_{\overline{q}}\otimes\hs_{X\setminus\overline{q}}$ and $\{|\psi_i\rangle\}$ is an orthonormal basis of $\hs_{\overline{q}}$. Second, let us write $\rightarrow^n$ for the composition of $n$ copies of $\rightarrow$. If $i_1\neq i_2$, then it is possible that $$(C_{i_1},|\theta_{i_1}\rangle)\rightarrow^{n_1}(\downarrow,|\theta_{i_1}^\prime\rangle)\ {\rm and}\ (C_{i_2},|\theta_{i_2}\rangle)\rightarrow^{n_2}(\downarrow,|\theta_{i_2}^\prime\rangle)$$ for $n_1\neq n_2$. This is why the reflexive and transitive closure $\rightarrow^\ast$ of $\rightarrow$ is used in the premise of the rule (QC). Finally, let $C\in\mathbf{QC}$ be a quantum circuit and $X\supseteq\mathit{qv}(C)$. Then the denotational semantics of $C$ over $X$ can be defined as operator $\llbracket C\rrbracket$ on $\hs_X$ as follows: $$\llbracket C\rrbracket|\varphi\rangle=|\psi\rangle\ {\rm if\ and\ only\ if}\ (C,|\varphi\rangle)\rightarrow^\ast(\downarrow,|\psi\rangle).$$ It is easy to show that 
\begin{equation}\label{qcase-3}\left\llbracket\mathbf{qif}[\overline{q}]\ \left(\square_{i=1}^d |\psi_i\rangle\rightarrow C_i\right)\ \mathbf{fiq}\right\rrbracket=\sum_{i=1}^d|\psi_i\rangle\langle\psi_i|\otimes\llbracket C_i\rrbracket.\end{equation} Furthermore, if we adopt the matrix representation of operators in the orthonormal basis $\{|\psi_i\rangle\}$, then (\ref{qcase-3}) can be written as the diagonal matrix $\mathit{diag}(\llbracket C_1\rrbracket,...,\llbracket C_d\rrbracket)$. This confirms that semantically, quantum case statement is exactly the same as quantum multiplexor \cite{Markov}. 

\subsection{Illustrative Examples}To illustrate their applicability, let us present several examples showing that the circuit constructs introduced in Definition \ref{def-circ}; in particular quantum case statement, can be used to define some commonly used quantum gates conveniently. 

\begin{exam}Assume that the single-qubit identity matrix $I$, the NOT matrix $X$, and $\overline{R}_x(\theta)=iR_x(2\theta)$ are unitary matrix constants in $\mathcal{U}$, where $R_x(\theta)$ is the rotation about the $x$ axis:  
$$R_x(\theta)=\left(\begin{array}{cc}i\cos\frac{\theta}{2} & \sin\frac{\theta}{2}\\ \sin\frac{\theta}{2} &i\cos\frac{\theta}{2}\end{array}\right)$$
 Then:\begin{enumerate}\item   
The CNOT (Controlled-NOT) gate with $q_1$ as its control qubit can be defined by
\begin{align*}\mathrm{CNOT}[q_1,q_2]:=\ &\mathbf{qif}[q_1]\ |0\rangle\rightarrow I[q_2]\\ &\quad\ \ \ \square\ |1\rangle\rightarrow X[q_2]\\ 
&\mathbf{fiq}
\end{align*} 
\item The Toffoli gate with $q_1,q_2$ as its control qubits is defined by 
\begin{align*}\mathrm{Toffoli}[q_1,q_2,q_3]:=\ &\mathbf{qif}[q_1,q_2]\ |00\rangle\rightarrow I[q_3]\\ &\qquad\quad\ \square\ |01\rangle\rightarrow I[q_3]\\ &\qquad\quad\ \square\ |10\rangle\rightarrow I[q_3]\\ &\qquad\quad\ \square\ |11\rangle\rightarrow X[q_3]\\ 
&\mathbf{fiq} 
\end{align*} 
\item The Deustch gate with $q_1,q_2$ as its control qubits is defined by 
\begin{align*}\mathrm{Deutsch}(\theta)[q_1,q_2,q_3]:=\ &\mathbf{qif}[q_1,q_2]\ |00\rangle\rightarrow I[q_3]\\ &\qquad\quad\ \square\ |01\rangle\rightarrow I[q_3]\\ &\qquad\quad\ \square\ |10\rangle\rightarrow I[q_3]\\ &\qquad\quad\ \square\ |11\rangle\rightarrow \overline{R}_x(\theta)[q_3]\\ 
&\mathbf{fiq} 
\end{align*} Note that $\mathrm{Deutsch}(\frac{\pi}{2})=\mathrm{Toffoli}$. 
\item The Fredkin gate with $q_1$ as its control qubit is defined by \begin{align*}\mathrm{Fredkin}[q_1,q_2,q_3]:=\ &\mathbf{qif}[q_1]\ |0\rangle\rightarrow I[q_2];I[q_3]\\ &\quad \ \ \ \square\ |1\rangle\rightarrow \mathrm{SWAP}[q_2,q_3]\\
&\mathbf{fiq} 
\end{align*} where the swap gate: \begin{align*}\mathrm{SWAP}[q_2,q_3]:=\mathrm{CNOT}[q_2,q_3];&\mathrm{CNOT}[q_3,q_2];\\ &\mathrm{CNOT}[q_2,q_3].\end{align*}
\end{enumerate}\end{exam}

Two quantum circuits $C_1,C_2\in\mathbf{QC}$ are said to be equivalent, written $C_1\equiv C_2$, if for any $|\psi\rangle,|\psi^\prime\rangle$, $$(C_1, |\psi\rangle)\rightarrow (\downarrow, |\psi^\prime\rangle)\ {\rm if\ and\ only\ if}\ (C_2,|\psi\rangle)\rightarrow(\downarrow, |\psi^\prime\rangle).$$ Then we have: 
\begin{exam}  It is easy to verify that 
\begin{align*}\mathbf{qif}[q_1] &|+\rangle\rightarrow I[q_2]\ \square\ |-\rangle\rightarrow Z[q_2]\ \mathbf{fiq}
\\ &\equiv\ \mathbf{qif}[q_2] |0\rangle\rightarrow I[q_1]\ \square\ |1\rangle\rightarrow X[q_2]\ \mathbf{fiq}.
\end{align*}
\end{exam}

\section{Quantum Circuits Defined with Classical Variables}\label{sec-c-variables}

To increase the expressive power, in this section we embed $\mathbf{QC}$ into a classical programming language. Our aim for this embedding is to allow us to use classical expressions as parameters in quantum recursive programs.  

\subsection{Syntax} For simplicity, let us choose classical \textbf{while}-language as the host language. Thus, we obtain: 

\begin{defn}\label{def-circ-1}Quantum circuits $C\in\mathbf{QC^+}$ with classical variables are defined by the syntax:
\begin{equation}\label{def-syntax-1}\begin{split}C::=\ \mathbf{skip}\ &|\ \overline{x}:=\overline{t}\ |\ U[\overline{q}]\ |\ C_1;C_2
\\ &|\ \mathbf{if}\ b\ \mathbf{then}\ C_1\ \mathbf{else}\ C_2\ \mathbf{fi}\\ &|\ \mathbf{while}\ b\ \mathbf{do}\ C\ \mathbf{od}
\\ &|\ \mathbf{qif}[\overline{q}]\ \left(\square_{i=1}^d |\psi_i\rangle\rightarrow C_i\right)\ \mathbf{fiq}\end{split}\end{equation}
where $\overline{x}$ is a string of classical simple or subscripted variables, $\overline{t}$ is a string of classical expression, $b$ is a Boolean expression, and other condition are the same as in Definition \ref{def-circ}.

The quantum variables $\mathit{qv}(C)$ in $C\in\mathbf{QC}^+$ are defined inductively as follows:\begin{enumerate} 
\item $\mathit{qv}(\mathbf{skip})=\mathit{qv}(\overline{x}:=\overline{t})=\emptyset$;
\item $\mathit{qv}(U[\overline{q}])=\overline{q}$;
\item $\mathit{qv}(C_1;C_2)=\mathit{qv}(\mathbf{if}\ b\ \mathbf{then}\ C_1\ \mathbf{else}\ C_2\ \mathbf{fi})=\mathit{qv}(C_1)\cup\mathit{qv}(C_2)$;
\item $\mathit{qv}(\mathbf{while}\ b\ \mathbf{do}\ C\ \mathbf{od})= \mathit{qv}(C)$;
\item $\mathit{qv}(\mathbf{qif}[\overline{q}]\ \left(\square_{i=1}^d |\psi_i\rangle\rightarrow C_i\right)\ \mathbf{fiq})=\overline{q}\cup\left(\bigcup_{i=1}^d \mathit{qv}(C_i)\right).$
\end{enumerate} \end{defn}

As usual in classical programming, a conditional of the form $\mathbf{if}\ b\ \mathbf{then}\ C_1\ \mathbf{else}\ \mathbf{skip}\ \mathbf{fi}$ will be simply written as $\mathbf{if}\ b\ \mathbf{then}\ C_1\ \ \mathbf{fi}$. 

\begin{rem}Our aim of introducing classical computation in $\mathbf{QC}^+$ is to enable quantum circuits be defined using classical expressions as their parameters. So, in a sense, the connection between classical and quantum variables in $\mathbf{QC}^+$ is unidirectional from classical ones to quantum ones. But a connection from quantum variables to classical ones can also be introduced by adding statements of the form $x:=M[\overline{q}]$, meaning that the outcome of measurement $M$ on quantum variables $\overline{q}$ is stored in classical variable $x$. For simplicity of presentation, however, we choose not to consider it in this paper.    
\end{rem}

\subsection{Operational semantics}

To define operational semantics of $\mathbf{QC}^+$, we need to modify the definition of configuration in order to accommodate classical variables. A configuration is now defined as a triple $(C,\sigma,|\psi\rangle)$, where $C\in\mathbf{QC}^+$ is a quantum variable with classical variables, $\sigma$ is a state of classical variables, and $|\psi\rangle$ is a pure quantum state in $\hs_X$ for some $\mathbf{qv}(C)\subseteq X\subseteq\mathit{QV}$. For simplicity, we abuse a bit of notation and still use $\mathcal{C}$ to denote the set of all configurations. 

\begin{defn}\label{def-sem-classical}The operational semantics of quantum circuits in $\mathbf{QC}^+$ is the transition relation $\rightarrow\ \subseteq\mathcal{C}\times\mathcal{C}$ between configurations defined by the transitional rules given in Table \ref{circuit-rules-1}.
\begin{table*}[t]
\begin{equation*}\begin{split}&({\rm SK})\ \ 
(\mathbf{skip},\sigma, |\psi\rangle)\rightarrow (\downarrow,\sigma, |\psi\rangle)\qquad\qquad\qquad\qquad\ \ \ \ \ \qquad ({\rm SC})\ \ \frac{(C_1,\sigma, |\psi\rangle)\rightarrow (
C_1^{\prime},\sigma^\prime, |\psi^{\prime}\rangle)}{(
C_1;C_2,\sigma, |\psi\rangle)\rightarrow (C_1^{\prime};C_2,\sigma^\prime, |\psi^\prime\rangle)}\\ 
&({\rm AS})\ \ (\overline{x}:=\overline{t},\sigma, |\psi\rangle)\rightarrow (\downarrow, \sigma[\overline{x}\leftarrow\sigma(\overline{t})], |\psi_\sigma\rangle)\qquad\qquad\qquad ({\rm GA})\ \ 
(U[\overline{q}], \sigma, |\psi\rangle)\rightarrow (
\downarrow,\sigma, (U\otimes I_{\mathit{QV}\setminus\overline{q}})|\psi\rangle)\\ 
&({\rm QC})\ \ \frac{|\psi\rangle=\sum_{i=1}^d\alpha_i|\psi_i\rangle_{\overline{q}}|\theta_i\rangle\qquad\qquad (C_i,\sigma, |\theta_i\rangle)\rightarrow^\ast(\downarrow,\sigma^\prime, |\theta_i^\prime\rangle)\ (i=1,...,n)}{\left(\mathbf{qif}[\overline{q}]\ \left(\square_{i=1}^d |\psi_i\rangle\rightarrow C_i\right)\ \mathbf{fiq},\sigma, |\psi\rangle\right)\rightarrow \left(\downarrow,\sigma^\prime, \sum_{i=1}^d\alpha_i|\psi_i\rangle_{\overline{q}}|\theta_i^\prime\rangle\right)}
\end{split}\end{equation*}
\caption{Transition Rules for Quantum Circuits with Classical Variables. In rule (AS), let $\overline{x}=x_1,...,x_n$ and $\overline{t}=t_1,...,t_n$. Then $\sigma(t_i)$ denotes the value of expression $t_i$ in state $\sigma$, and $\sigma[\overline{x}\leftarrow\sigma(\overline{t})]$ is the state of classical variables obtained by replacing the value of $x_i$ in $\sigma$ with $\sigma(t_i)$ simultaneously for all $1\leq i\leq n$.}\label{circuit-rules-1}
\end{table*}
\end{defn}

The rules (SK), (SC), (AS) and (GA) are easy to understand. One design decision in the rule (QC) needs an explanation. In its premise, when starting in the state classical state $\sigma$, the executions of all branches $(C_i,\sigma,|\theta_i\rangle)$ $(i=1,...,n)$ are required to terminate in the same classical state $\sigma^\prime$. At the first glance, this is a very strong requirement and hard to meet in practical applications. Indeed, as we will see in the next subsection and Example \ref{exam-control}, it can be easily achieved by introducing local variables. 

\subsection{Local variables} 

In this subsection, we further introduce local classical variables into $\mathbf{QC}^+$ by extending its syntax with the following clause:
\begin{equation}\label{def-block}C::=\mathbf{begin\ local}\ \overline{x}:=\overline{t};C\ \mathbf{end}\end{equation} where $\overline{x}$ is a sequence of classical variables and $\overline{t}$ a sequence of classical expressions. A statement of the form (\ref{def-block}) is called a block statement, and its operational semantics is defined by the rule (BS) in Table \ref{circuit-rules-1+}. 
\begin{table*}[t]
\begin{equation*}({\rm BS})\ \ (\mathbf{begin\ local}\ \overline{x}:=\overline{t},C\ \mathbf{end}, \sigma, |\psi\rangle)\rightarrow (\overline{x}:=\overline{t};C;\overline{x}:=\sigma(\overline{x}),\sigma,|\psi\rangle)
\end{equation*}
\caption{Transition Rule for Local Variables.}\label{circuit-rules-1+}
\end{table*}

The rule (BS) is very similar to the rule defining the operational semantics of local variables in classical programs (see for example \cite{Apt09}, the rule (ix) on page 154). 
In the execution of block statement (\ref{def-block}) starting in classical state $\sigma$, the \textit{local variables} $\overline{x}$ are first initialised by assignment $\overline{x}:=\overline{t}$, then the circuit $C$ within the statement is executed. After that, $\overline{x}$ resume their original values in $\sigma$. It is easy to see that if $\overline{x}$ is an empty sequence, then the block statement can be identified with circuit $C$ within it.  

We will see in Example \ref{exam-control} how local variables can help in describing quantum recursive programs. 

\section{Quantum recursive programs without parameters}\label{sec-non-parameters}

Now we are ready to define a language $\mathbf{RQC}^+$ of recursively defined quantum circuits. In this section, we only consider quantum recursive circuits without parameters. More general quantum recursive circuits with parameters will be considered in the next section. As we will see shortly, at the level of syntax, quantum recursive circuits and classical recursive programs (see for example \cite{Apt09}, Chapters 4 and 5) are similar to each other. The major difference between them appears at the level of semantics, where quantum case statements involved in the former will exhibit superposition of the executions of multiple circuits within recursive procedures.  

\subsection{Syntax}Let us first define the syntax of $\mathbf{RQC}^+$. We add a set of \textit{procedure identifiers}, ranged over by symbols $P, P_1,P_2,...$ into the alphabet of $\mathbf{QC}^+$. Then the syntax of $\mathbf{RQC}^+$ is defined by extending the syntax (\ref{def-syntax-1}) and (\ref{def-block}) of $\mathbf{QC}^+$ by adding the clause:
\begin{equation}\label{def-syntax-2}C::=P\end{equation} with quantum variables $\mathit{qv}(P)=\emptyset$. As in classical recursive programming \cite{Apt09}, an occurrence of a procedure identifier in a program is called a \textit{procedure call}. We assume that each procedure identifier $P$ is defined by a \textit{declaration} of the form \begin{equation}\label{def-sem-2}P\Leftarrow C\end{equation} where $C\in\mathbf{RQC}^+$ is called the procedure body. Note that in the declaration (\ref{def-sem-2}), $P$ may appear in the procedure body $C$; the occurrences of $P$ in $C$ are thus called \textit{recursive calls}. We assume a fixed set $\mathcal{D}$ of procedure declarations.  

\subsection{Operational semantics}To define the operational semantics of $\mathbf{RQC}^+$, we first generalise the notion of configuration $(C,\sigma, |\psi\rangle)$ by allowing $C\in\mathbf{RQC}^+$. Then the operational semantics is the transition relation $\rightarrow$ between configurations defined by the transition rules given in Tables \ref{circuit-rules-1} and \ref{circuit-rules-1+} together with the following \textit{copy rule}: 
\begin{equation}\label{copy-rule}({\rm CR})\qquad \frac{P\Leftarrow C\in\mathcal{D}}{(P,\sigma, |\psi\rangle)\rightarrow (C,\sigma, |\psi\rangle)}\end{equation}
Intuitively, the copy rule allows that a procedure call is dynamically replaced by the procedure body of its declaration given in declarations $\mathcal{D}$. 

\subsection{Illustrative Examples}
The following two examples show how some large unitary transformations (i.e quantum gates) can be elegantly described as quantum recursive programs defined above. 

\subsubsection{Controlled unitary transformations} Controlled unitaries are a class of quantum gates widely used in quantum computing. Mathematically, let $U$ be a unitary operator on a single qubit and $n$ a positive integer. Then the controlled-$U$ gate $C^{(n)}(U)$ with $q_1,...,q_n$ as its control qubits and $q_{n+1}$ as its target qubit is defined by \begin{align*}&C^{(n)}(U)|i_1,...,i_n\rangle|\psi\rangle\\ &\qquad =\begin{cases}
|i_1,...,i_n\rangle U|\psi\rangle\ &{\rm if}\ i_1=...=i_n=1;\\ |i_1,...,i_n\rangle|\psi\rangle &{\rm otherwise}
\end{cases}\end{align*} for any $i_1,...,i_n\in\{0,1\}$ and $|\psi\rangle\in\hs_2$. 

Using a quantum programming language without recursion, one has to define $C^{(n)}$ for different integers $n$ individually. 
Now within the recursion scheme introduced above, we can define $C^{(n)}$ in a uniform way:
\begin{exam}\label{exam-control} Let $q$ be a qubit array of type $\mathbf{integer}\rightarrow\hs_2$. Then the controlled-$U$ gate on the section $q[\mathit{first}:\mathit{last}]$ with the first $(\mathit{last}-\mathit{first})$ qubits as its control qubits and the last one as its target qubit can be written as the recursive program:
\begin{equation*}\begin{split}&C^{(\ast)}(U)\ \Leftarrow\\ &\mathbf{if}\ \mathit{first}=\mathit{last}\\ &\ \ \ \mathbf{then}\ U[q[\mathit{last}]]\\
&\ \ \ \mathbf{else}\ \mathbf{qif}[q[\mathit{first}]]|0\rangle\rightarrow\mathbf{skip}\\ 
&\qquad\qquad\qquad\quad\square\ |1\rangle\rightarrow \mathbf{begin\ local}\ \mathit{first}:=\mathit{first}+1;\\ &\qquad\qquad\qquad\qquad\qquad\quad C^{(\ast)}(U)\ \mathbf{end}\\ &\qquad\quad\ \mathbf{fiq}\\ 
&\mathbf{fi}
\end{split}\end{equation*}    
\end{exam}

It is worth noting that a block statement with local variables is employed in the above program $C^{(\ast)}(U)$ to guarantee that different branches of a quantum case statement terminate in the same classical state (see the explanation of the transition rule (QC) given after Definition \ref{def-sem-classical}).     

\subsubsection{Quantum Fourier transforms} As a key subroutine, quantum Fourier transforms appear in many important quantum algorithms, including Shor's factoring algorithm. The quantum Fourier transform $\mathit{QFT}(n)$ on $n$ qubits is mathematically defined by 
\begin{equation}\label{QFT}\mathit{QFT}(n)|j\rangle=\frac{1}{\sqrt{2^n}}\sum_{k=0}^n e^{2\pi ijk/2^n}|k\rangle
\end{equation} for $j=0,1,...,2^n -1.$ If we use the binary representation $j=j_1j_2...j_n=\sum_{l=1}^n j_l 2^{n-l}$ and binary fraction $0.k_1k_2...k_m=\sum_{l=1}^mk_l 2^{-l},$ then the defining equation (\ref{QFT}) of $\mathit{QFT}(n)$ can be rewritten as
\begin{equation}\label{QFT-1}\mathit{QFT}(n)|j_1,...,k_n\rangle=\frac{1}{\sqrt{2^n}}\bigotimes_{l=1}^n\left(|0\rangle+e^{2\pi i 0.j_{n-l+1}...j_n}|1\rangle\right).
\end{equation} As shown in Table \ref{QFT-table}, $\mathit{QFT}(n)$ can be decomposed into a sequence of single-qubit and two-qubit basic gates, namely the Hadamard gate $H$ and controlled-rotations $C(R_l)$ $(l=1,...,n)$, where $$R_l=\left(\begin{array}{cc}1&0\\ 0&e^{2\pi i/2^l}\end{array}\right)$$ is a single-qubit gate. 
\begin{table*}[t]
\begin{align*}\mathit{QFT}(n)[q_1,...,q_n]::=\ &H[q_1]; C(R_2)[q_2,q_1];C(R_3)[q_3,q_1];...;C(R_{n-1})[q_{n-1},q_1];C(R_n)[q_n,q_1];\\
&H[q_2];C(R_2)[q_3,q_2];C(R_3)[q_4,q_2];...;C(R_{n-1})[q_n,q_2];\\
&\qquad\qquad\qquad\qquad ..............................\\
&H[q_{n-1}];C(R_2)[q_n,q_{n-1}];\\
&H[q_n]; \\
&\mathit{Reverse}[q_1,...,q_n]
\end{align*}
\caption{A quantum circuit for quantum Fourier transform. Here,  $C(R_l)[q_i,q_j]$ stands for the controlled-$R_l$ with $q_i$ as its control qubit and $q_j$ as its target qubit, and $\mathit{Reverse}[q_1,...,q_n]$ is the quantum gate that reverses the order of qubits $q_1,...,q_n$.}\label{QFT-table}
\end{table*}

In a quantum programming language that does not support recursion, one has to program $\mathit{QFT}(n)$ in a way similar to Table \ref{QFT-table}. For a large number $n$ of qubits, the size of such a $\mathit{QFT}(n)$ program will be very large. Using recursion, however, we can write $\mathit{QFT}(n)$ as a program, of which the size is independent of the number $n$ of qubits:   

\begin{exam} Let $q$ be a qubit array of type $\mathbf{integer}\rightarrow\hs_2$. Then quantum Fourier transform on the section $q[m,n]$ can be written as the following recursive program:
\begin{align*}
\mathit{QFT}(m,n)\ \Leftarrow\ H[q[m]];\ &\mathbf{if}\ m<n\ \mathbf{then}\ \mathit{Rotate}(m,n);\\
&\qquad\qquad\qquad\mathit{QFT}[m+1,n]\\
&\mathbf{fi};\\
&\mathit{Reverse}(m,n)
\end{align*}
\begin{align*}\mathit{Rotate}(m,n)\ \Leftarrow \ &\mathit{Rotate}(m,n-1);\\ &\mathbf{qif}[q[n]]|0\rangle\rightarrow\mathbf{skip}\\
&\ \ \ \ \quad\square\ |1\rangle\rightarrow R_n[q[m]]\\
&\mathbf{fiq}
\end{align*}
\begin{align*}\mathit{Reverse}(m,n)\ \Leftarrow\ &\mathbf{if}\ m<n\ \mathbf{then}\ \mathit{SWAP}[q[m],q[n]];\\ &\qquad \mathbf{if}\ m+2\leq n\ \mathbf{then}\\ &\qquad\qquad\mathit{Reverse}(m+1,n-1)\\ &\qquad\mathbf{fi}
\\ &\mathbf{fi}
\end{align*}
\end{exam}

\section{Quantum recursive programs with parameters}\label{sec-parameters}

In this section, we further expand the language $\mathbf{RQC}^+$ defined in the last section to $\mathbf{RQC}^{++}$ of quantum recursive circuits with \textit{classical} parameters.

\subsection{Syntax} We add a set of procedure identifiers $P, P_1,P_2,...$ into the alphabet of $\mathbf{QC}^{+}$. Each identifier $P$ is given an arity $\mathit{ar}(P)$. 
Then the syntax of $\mathbf{RQC}^{++}$ is defined by the syntax (\ref{def-syntax-1}) and (\ref{def-block}) of $\mathbf{QC}^+$ together with the following clause: 
\begin{align}\label{def-syntax-2}C::=\ & P(t_1,...,t_n)\end{align}
The above syntax of $\mathbf{RQC}^{++}$ is very similar to that of classical recursive programs with parameters given in \cite{Apt09}. In procedure call (\ref{def-syntax-2}), $P$ is a procedure identifier with $\mathit{ar}(P)=n$, and $t_1,...,t_n$ are classical expressions, called \textit{actual parameters}. Whenever $n=0$, procedure $P(t_1,...,t_n)$ degenerates to a procedure without parameters considered in the last section. 

Each procedure identifier $P$ is defined by a declaration of the form: \begin{equation}P(u_1,...,u_n)\Leftarrow C\end{equation} where $u_1,...,u_n$ are classical simple variables, called \textit{formal parameters}; and the procedure body $C\in\mathbf{RQC}^{++}.$ We assume a fixed set $\mathcal{D}$ of procedure declarations.  

\subsection{Operational semantics} 
Now configurations are triples $(C,\sigma,|\psi\rangle)$ with $C\in\mathbf{RQC}^{++}$ and $\sigma, |\psi\rangle$ being as in Section \ref{sec-non-parameters}. The semantics of $\mathbf{RQC}^{++}$ is then a transition relation between configurations defined by the rules in Tables \ref{circuit-rules-1} and \ref{circuit-rules-1+} together with the recursive rule (RC) in Table \ref{circuit-rules-2}. 
\begin{table*}[t]
\begin{equation*}({\rm RC})\ \ \frac{P(u_1,...,u_n)\Leftarrow C\in\mathcal{D}}{(P(t_1,...,t_n),\sigma, |\psi\rangle)\rightarrow (\mathbf{begin\ local}\ \overline{u}:=\overline{t};C\ \mathbf{end},\sigma, |\psi\rangle)}
 \end{equation*}
\caption{Transition Rule for Quantum Recursive Circuits with Parameters.}\label{circuit-rules-2}
\end{table*} 

\subsection{Illustrative examples}

\subsubsection{Controlled unitary transformations revisited} Using recursion with parameters introduced in this section, controlled unitaries can be programmed in a more compact way than Example \ref{exam-control}: 
\begin{exam}The controlled-$U$ gate on the section $q[m:n]$ with the first $(n-m)$ qubits as its control qubits and the last one as its target qubit can be written as the recursive program:
\begin{equation*}\begin{split}C^{(\ast)}(U)(m,n)\ \Leftarrow\ &\mathbf{if}\ m=n\\ &\ \ \ \mathbf{then}\ U[q[n]]\\
&\ \ \ \mathbf{else}\ \mathbf{qif}[q[m]]|0\rangle\rightarrow\mathbf{skip}\\ 
&\qquad\qquad\qquad\square\ |1\rangle\rightarrow C^{(\ast)}(U)[m+1,n]\\ &\qquad\quad\ \mathbf{fiq}\\ 
&\mathbf{fi}
\end{split}\end{equation*}    
\end{exam}
In particular, it is interesting to note that different from Example \ref{exam-control}, the $C^{(\ast)}(U)$ program in the above example does not use any block statement and local variable.   

\begin{rem}\label{multi-coin}The idea of the above example can be easily generalised to give a recursive definition of a quantum case statement with multiple quantum coins of the form:
\begin{equation}\label{multi-coin1}\mathbf{qif}[q[1:k]](\square_{x\in\{0,1\}^k}|x\rangle\rightarrow U_x[q[k+1]])\ \mathbf{fiq}
\end{equation} in terms of quantum case statements with a single quantum coin of which the number of branches is fixed. 
\end{rem}

\subsubsection{Quantum state preparation} 
Quantum state preparation (QSP) is a basic procedure employed in many quantum algorithms; in particular, in quantum simulation and quantum machine learning.  
The problem is as follows. Given an $N$-dimensional complex vector $\mathbf{a}=\left(a_j\right)_{j=0}^{N-1}\in\mathbb{C}^N,$  where $N=2^n$. Our goal is to generate the $n$-qubit state:
$$\frac{1}{\sqrt{a}}\sum_{j=0}^{N-1}\sqrt{a_j}|j\rangle$$ from the basis state $|0\rangle^n$, where $a=\sum_{j=0}^{N-1}|a_j|.$ 

For each $0\leq j<N$, and for any $0\leq l<r \leq N$, define $\theta_j$ and $S_{l,r}$ such that $a_j=e^{i\theta_j}|a_j|$ and $S_{l,r}=\sum_{j=l}^{r-1}|a_j|.$ Then the QSP algorithm (which slightly generalises the one in \cite{KP17}) consists of $n$ steps. For $0\leq k< n$, in the $k$th step, it performs the transformation:
\begin{equation*} 
\begin{cases}
|0\rangle^ n\mapsto U_{0,0}|0\rangle |0\rangle^{\otimes (n-1)}& k=0;\\
|x\rangle|0\rangle^{\otimes (n-k)} \mapsto |x\rangle U_{k,x}|0\rangle |0\rangle^{\otimes(n-k-1)}& 1\leq k<n
\end{cases}
\end{equation*} for all $0\leq x<2^k-1$,
where $U_{k,x}$ is a single qubit gate such that:
$$U_{k,x}|0\rangle=\sqrt{\gamma_x}|0\rangle+e^{i\beta_x/2}\sqrt{1-\gamma_x}|1\rangle,$$
and $\gamma_x=\frac{S_{u,w}}{S_{u,v}}, \beta_x=\theta_w-\theta_u, u=2^{n-k}x,\quad v=2^{n-k}x+2^{n-k}$ and $w=\frac{u+v}{2}.$ Using the language $\mathbf{RQC}^+$, the QSP algorithm can be elegantly rewritten as a quantum recursive program:
\begin{exam}Let 
\begin{align*}\mathit{QSP}(k,n)\ \Leftarrow
\ &\mathbf{if}\ k=0\ \mathbf{then}\ U_{0,0}[q[1]]\ \mathbf{else}\\ 
&\ \ \mathbf{if}\ 1\leq k<n\ \mathbf{then}\\
&\ \ \ \ \mathbf{qif}[q[1:k]](\square |x\rangle\rightarrow U_{k,x}[q[k+1]];\\ &\qquad\qquad\qquad\qquad\qquad \mathit{QSP}(k+1,n))\\
&\ \ \ \ \mathbf{fiq}\\
&\ \ \mathbf{fi}\\
&\mathbf{fi}
\end{align*} Then one calls $\mathit{QSP}(0,n)$ for the quantum state preparation. 
\end{exam}

At the first glance, there seems a bug in the above program: the number $2^k$ of branches of the $\mathbf{qif}$-statement varies as the number $k$ of coin qubits $q[1:k]$. But it is actually not a bug because as pointed out in Remark \ref{multi-coin}, the $\mathbf{qif}$-statement with coins $q[1:k]$ can be recursively defined in terms of $\mathbf{qif}$-statement with a single coin qubit.    

\subsubsection{Quantum Random-Access Memory (QRAM)} Many applications of quantum computing (from optimisation and machine learning to cryptanalysis) presume the existence of QRAM, a quantum counterpart of RAM (Random-Access Memory) in classical computing~\cite{Jaq23}. The strongest type of QRAM is called QRAQM (Quantum Random-Access Quantum Memory), which stores quantum data and access data based on addresses that are themselves a quantum state in a superposition. Among several equivalent forms, we consider a QRAQM that performs the following transformation: for any data set $D[0:N]$ with $N=2^n -1$, and for any address $0\leq j \leq N$, 
\begin{equation}
    |j\rangle |D[0:N]\rangle \mapsto |j\rangle |D[j]\rangle |D[0:j-1]\rangle |D[j+1:N]\rangle.
    \label{eq:QRAM}
\end{equation}
Intuitively, given an address $j$, the desired data element $D[j]$ is swapped out. A simple (but not very efficient) implementation of QRAQM can be written as a quantum recursive program:
\begin{exam}\label{exam-qram} We use $q_a[1:n]$ for the address register holding $|j\rangle$ 
and $q_D[0:N]$ for the data register holding $|D[0:N]\rangle$ on the LHS of \eqref{eq:QRAM}. Let
\begin{equation*}\begin{split}
    U(l,r,k)\Leftarrow \ &\mathbf{if}\ k\leq n \ \mathbf{then}\\ 
    &\ \ \ \mathbf{begin\ local}\ m:=\lfloor(l+r)/2\rfloor;\\
    &\ \ \ \ \ \ \mathbf{qif}[q_a[k]]\ |0\rangle\rightarrow U(l,m,k+1)\\
    &\ \ \ \ \ \qquad\quad\ \square\ \ |1\rangle\rightarrow U(m+1,r,k+1);\\
    &\ \ \ \ \quad\quad\quad\quad\ \ \ \mathit{SWAP}[q_D[l],q_D[m+1]]\\
    &\ \ \ \ \ \ \mathbf{fiq}\\
    &\ \ \ \mathbf{end}\\ 
    &\mathbf{fi}
\end{split}\end{equation*}
Then one calls $U(0,N,1)$ for the QRAQM operation.\end{exam}

It is interesting to note that a Divide-and-Conquer strategy was employed in the above example where QRAM is divided into two subproblems smaller than the original one that are then solved respectively in each of the branches of a quantum case statement. 

\section{Conclusion}

This \textit{short paper} introduces a new scheme of quantum recursive programming. The basic ideas of this scheme of quantum recursion are illustrated through a series of interesting examples. It should be emphasised that this scheme of quantum recursion is defined based on the notion of quantum case statement; indeed, it cannot be realised without quantum case statements, as we can observe from the examples. 

This paper is merely one of the first steps toward a theory of quantum recursive programming. Plenty of problems about quantum recursions remain unsolved. Here, we would like to mention the following two open problems:   
\begin{itemize}
\item \textit{Implementation of quantum recursion}: Classical recursive programs are usually implemented employing stack \cite{Dijk, SICP}. How can we implement the kind of quantum recursion introduced in this paper? Indeed, a notion of quantum stack has still not been properly defined. 
\item \textit{More sophisticated quantum recursive programming techniques}: A simple Divide-and-Conquer strategy was employed in Example \ref{exam-qram}. 
Several other quantum Divide-and-Conquer strategies have been proposed in the literature (see for example \cite{Childs}). It is interesting to see whether or not and how quantum algorithms developed with these Divide-and-Conquer strategies can be recursively programmed.  Furthermore, how can quantum recursive programming be combined with structural development techniques of quantum algorithms as recently proposed in \cite{Rossi23}. 
\end{itemize}

\section*{Acknowledgments}This work was partly supported by the National Natural Science Foundation of China (Grant No: 61832015).
 Zhicheng Zhang was supported by the Sydney Quantum Academy, NSW, Australia.

\end{document}